\documentclass[aps,pre,twocolumn,notitlepage,superscriptaddress]{revtex4-1}

\usepackage{amsmath}
\usepackage{amssymb}
\usepackage{hyperref}
\usepackage{graphicx}
\usepackage[arrow, matrix, curve]{xy}
\usepackage{bbm}
\usepackage{bm}
\usepackage{yhmath}
\usepackage{color}
\usepackage{physics}
\usepackage{bbold}

\usepackage{color}

\renewcommand\vec[1]{\boldsymbol{\mathrm{#1}}}

\newcommand\diff{\mathrm{d}}

\usepackage[usenames,dvipsnames]{xcolor}
\hypersetup{colorlinks=true, linkcolor=BrickRed, urlcolor=blue!50!black, citecolor=blue!50!black}

\makeatletter
\newcommand\hide@visible[1]{%
  \bgroup\fboxsep=.3ex\colorbox{Gray}{begin hide}%
  #1\colorbox{Gray}{end hide}\egroup%
}
\newcommand\hide@hidden[1]{%
  \bgroup\fboxsep=.3ex\colorbox{Gray}{hidden text}%
}
\newcommand\hide@invisible[1]{}
\newcommand\makevisible{\let\hide\hide@visible}
\newcommand\makehidden{\let\hide\hide@hidden}
\newcommand\makeinvisible{\let\hide\hide@invisible}
\makeatother
\makehidden

\begin{document}

\title{Two-Dimensional Active Brownian Particles Crossing a Parabolic Barrier: \\ Transition-Path Times, Survival Probability, and First-Passage Time.}

\author{Michele Caraglio}
\email{Michele.Caraglio@uibk.ac.at}
\affiliation{Institut f\"ur Theoretische Physik, Universit\"at Innsbruck, Technikerstra{\ss}e 21A, A-6020, Innsbruck, Austria}

\date{\today}

\begin{abstract}
We derive an analytical expression for the propagator and the transition path time distribution of a two-dimensional active Brownian particle crossing a parabolic barrier with absorbing boundary conditions at both sides.
By taking those of a passive Brownian particle as basis states and dealing with the activity as a perturbation, our solution is expressed in terms of the perturbed eigenfunctions and eigenvalues of the associated Fokker-Planck equation once the latter is reduced by taking into account only the coordinate along the direction of the barrier and the self-propulsion angle.
We show that transition path times are typically shortened by the self-propulsion of the particle.
Our solution also allows us to obtain the survival probability and the first-passage times distribution, which display a strong dependence on the particle's activity, while the rotational diffusivity influences them to a minor extent.
\end{abstract}

\maketitle

\section*{Introduction}

Active particles are characterized by the ability to consume energy from their environment in order to generate self-propelled directed motion~\cite{Bechinger2016,Marchetti2013,Romanczuk2012,Elgeti2015}.
Recent decades have seen significant and growing effort in investigating their dynamics because of their relevance in a wide variety of fields including biology~\cite{Needleman2017,Lipowsky2005}, biomedicine~\cite{Wang2012,Henkes2020}, robotics~\cite{Cheang2014,Erkoc2019}, social transport~\cite{Helbing2001}, and statistical physics~\cite{Cates2012,Chaudhuri2014,Fodor2016,Falasco2016,Speck2016,Fodor2018,Caraglio2022}.

For systems of interacting active particles novel emerging collective behavior arises~\cite{Kumar2014,Fily2012,Slowman2016,Stenhammar2015}.
Furthermore, even at the single-particle level, self-propelled particles show remarkable features such as, for example, accumulation near confining boundaries~\cite{Elgeti2013,Volpe2014}, non-Boltzmann stationary distributions~\cite{Tailleur2009,Malakar2018,Wagner2017,Malakar2020}, and oscillating intermediate scattering functions~\cite{Kurzthaler2016,Kurzthaler2017,Kurzthaler2018}.
However, notwithstanding the tremendous progress that has been achieved in this research field in the past decade, exactly solvable models, allowing a deeper understanding of some basic theoretical aspects, remain rare.
Exceptions include active Brownian particles (ABPs) in channels~\cite{Wagner2017} or sedimenting in a gravitational field~\cite{Hermann2018} and run-and-tumble particles in one dimension~\cite{Schnitzer1993,Tailleur2008,Tailleur2009,Malakar2018}.
The analytical solution of the time-dependent probability distribution of run-and-tumble and ABPs in free space is known only in the Fourier domain~\cite{Kurzthaler2016,Kurzthaler2017,Kurzthaler2018,Martens2012,Sevilla2015}.
More recently, the steady-state distribution of a two-dimensional ABP trapped in an isotropic harmonic potential has been derived~\cite{Malakar2020} and its time-dependent Fokker-Planck equation has been fully solved~\cite{Caraglio2022}.
Finally, the Fokker-Planck equation has been solved also for an ABP exploring a circular region with an absorbing boundary~\cite{DiTrapani2023} and, by disregarding one of the two spatial coordinates, for an ABP reaching an absorbing wall in two dimensions~\cite{Bauche2025}.

Here, we extend the set of exactly solvable models by considering an ABP navigating in a domain characterized a parabolic barrier along one direction and presenting absorbing boundaries at both sides of the barrier.
Building upon the theory developed in Refs.~\cite{Caraglio2022,DiTrapani2023}, we show that also in the current system a formally exact series expression for the probability propagator associated with the Fokker-Planck equation can be obtained once the basis states of a reference standard Brownian particle are known.
For the considered environment, we attain the basis states of a passive Brownian particle by extending to two dimensions the solution of the Fokker-Planck equation for a one-dimensional particle crossing a parabolic barrier as achieved in Ref.~\cite{Caraglio2018}.
In the present manuscript, we focus on the case of a domain infinitely extended in the direction perpendicular to the energy barrier and we found an expression for the reduced propagator taking into account the position along the direction of the barrier and the self-propulsion angle but not the position along the perpendicular direction.

Furthermore, again taking inspiration from Ref.~\cite{Caraglio2018}, we also derive an expression for the transition-path-times (TPTs) distribution.
In a barrier-crossing process, transition paths are defined as those trajectories originating from a point at one side of the barrier and ending at another point at the opposite side, without recrossing the initial position.
In the context of passive dynamics, both experimentalists~\cite{Chung2009,Neupane2016,Neupane2017,Neupane2018} and theorists~\cite{Zhang2007,Faccioli2012,Kim2015,Makarov2015,Laleman2017,Caraglio2020,Satija2020} devoted quite some attention to TPTs, because they carry important information about the reactive dynamics.

Finally, our solution allows us, by properly integrating the propagator, to find also the survival probability and the first-passage-time distribution~\cite{Redner2001,Metzler2013,Risken1989,Palyulin2019}.
The latter observables characterize several processes in nature~\cite{Gerstner1997,Sazuka2009,Baldovin2015,Chmeliov2013,Biswas2016} and plays a pivotal role in understanding transport properties and escape dynamics of living micro-organisms or artificial nano- and micro-particles in different environments~\cite{Redner2001}.

\section*{Model}

The stochastic overdamped motion of a two-dimensional ABP is completely characterized in terms of the propagator $\mathbb{P}(\vec{r}, \vartheta, t | \vec{r}_0 , \vartheta_0)$ which is the probability to find the particle at position $\vec{r}=(x,y)$ and orientation $\vartheta$ at lag time $t$ given the initial position $\vec{r}_0$ and orientation $\vartheta_0$ at time $t=0$.

In presence of a parabolic barrier along the $x$-direction, $U(x) = -k x^2/2$, with spring constant $k > 0$, the Fokker-Planck equation reads
\begin{align} \label{eq:eom_propagator_v0}
    \partial_t\mathbb{P}  = \Omega \mathbb{P} := &  D \partial_x \left( e^{\mu k x^2 / 2D} \partial_x e^{-\mu k x^2 / 2D} \mathbb{P} \right)  \nonumber \\ 
    & + D \partial_y^2 \mathbb{P} + D_{\text{rot}} \partial_\vartheta^2 \mathbb{P} -  v \vec{u} \cdot \vec{\nabla} \mathbb{P} \; ,
\end{align}
where $D$ and $D_{\text{rot}}$ are the translational and rotational diffusion coefficient, respectively, whereas $\mu$ is the mobility of the particle.
The ratio $D/\mu = k_B T$ defines the energy unit and introduces an effective temperature that for a passive particle corresponds to the temperature of the bath.
Finally, the particle is endowed with a self-propulsion having fixed velocity $v$ along the orientation $\vec{u} = (\cos \vartheta, \sin \vartheta)$.
The previous equation readily provides the formal solution 
\begin{align} \label{eq:formal_solution_v0} \mathbb{P}(\vec{r}, \vartheta, t | \vec{r}_0 , \vartheta_0) = e^{\Omega t} \delta(\vec{r}-\vec{r}_0) \delta(\vartheta-\vartheta_0) \; ,
\end{align}
of the propagator given the initial condition
\begin{align} \label{eq:initial_condition_v0}
    \mathbb{P}(\vec{r}, \vartheta, t=0 | \vec{r}_0 , \vartheta_0) =  \delta(\vec{r}-\vec{r}_0) \delta(\vartheta-\vartheta_0) \; .
\end{align}

Here, we are interested in investigating the dynamics of an ABP conditioned to the presence of absorbing boundaries at $x = \pm d$.
\begin{align} 
	\mathbb{P}(x=\pm d, y, \vartheta, t | \vec{r}_0 , \vartheta_0) = 0 \; , \label{eq:boundary_condition_x_v0}
\end{align}
and the further requirement that the initial position $\vec{r}_0$ is chosen in between the boundaries ($-d < x_0 < d$).

While absorbing boundaries are beneficial to develop a theoretical framework aiming at characterizing first-passage properties~\cite{Redner2001}, they also represent experimental or natural scenarios where particles are permanently removed upon reaching certain regions, such as chemical reactions at reactive surfaces or particles escaping confinement~\cite{Bressloff2023}.
Absorbing boundaries may also be considered as a limiting case of sticky boundary conditions, which are a way to formulate the accumulation process at conﬁning boundaries typically exhibited by active particles~\cite{Bressloff2023b}.
Finally and most importantly, absorbing boundaries are necessary to account for a correct definition of the transition-path time~\cite{Caraglio2018}:
Typically, the transition-path time refers to the time it takes for a system to cross from one stable state to another, without returning to the initial state. 
It captures only the duration of the actual transition, excluding the long waiting times the system might spend in the initial or final states.
In such a context, our absorbing boundaries could also be interpreted as fictitious surfaces separating the transition region, $-d<x<d$, from two arbitrarily defined potential basins.

Without any further condition restricting the movement of the particle along the $y$ direction, we are unfortunately unable to find an expression of the full propagator.
However, thanks to the fact that the diffusion processes along the $x$ and $y$ coordinates are decoupled and that the drift dynamics depends only on the self-propulsion direction, we can still find a solution for the reduced propagator $\widetilde{\mathbb{P}}(x, \vartheta, t | x_0 , \vartheta_0)$ which is the probability to find the particle at coordinate $x$ and orientation $\vartheta$ at lag time $t$ given the initial coordinate $x_0$ and orientation $\vartheta_0$ at time $t=0$.
For such a reduced propagator, Eq.~\eqref{eq:eom_propagator} reads
\begin{align} \label{eq:eom_propagator}
    \partial_t \widetilde{\mathbb{P}}  = \widetilde{\Omega} \widetilde{\mathbb{P}} := &  D \partial_x \left( e^{\mu k x^2 / 2D} \partial_x e^{-\mu k x^2 / 2D} \widetilde{\mathbb{P}} \right)  \nonumber \\ 
    &  + D_{\text{rot}} \partial_\vartheta^2 \widetilde{\mathbb{P}} -  v \cos \vartheta \, \partial_x \widetilde{\mathbb{P}} \; ,
\end{align}
and has a formal solution given by
\begin{align} \label{eq:formal_solution} 
	\widetilde{\mathbb{P}}(x, \vartheta, t | x_0 , \vartheta_0) = e^{\widetilde{\Omega} t} \delta(x-x_0) \delta(\vartheta-\vartheta_0) \; ,
\end{align}
with initial and boundary conditions respectively give by
\begin{align} \label{eq:initial_condition}
    \widetilde{\mathbb{P}}(x, \vartheta, t=0 | x_0 , \vartheta_0) =  \delta(x-x_0) \delta(\vartheta-\vartheta_0) \; .
\end{align}
and
\begin{align} 
	\widetilde{\mathbb{P}}(x=\pm d, \vartheta, t | x_0 , \vartheta_0) = 0 \; , \label{eq:boundary_condition_x}
\end{align}

In the rest of this manuscript, we will always refer to the above-stated reduced problem but, for the sake of notation simplicity, we will drop the $\widetilde{\bullet}$ symbol.

We exploit the distance $d$ to fix the length unit of the problem.
Taking the passive Brownian particle ($v=0$) as a reference, it is convenient to define the time unit $\tau$ as the typical time required by such a particle to cover the distance between the two boundaries along the $x$ direction, $\tau := d^2/D$.
The dynamics of the ABP is then described only by two independent dimensionless parameters: The P\'eclet number, $\text{Pe} := v \tau /d$, assessing the importance of the self-propulsion with respect to the diffusive motion, and the \textit{``rotationality''}, $\gamma :=  D_{\text{rot}} \tau$,  measuring the magnitude of the rotational diffusion.
The fraction $\text{Pe}/\gamma$ then describes the persistence of the particle's trajectories in free space.

In order to find an expression for the propagator that is a solution of Eq.~\eqref{eq:eom_propagator}, first we make a time-separation ansatz for the propagator, $\mathbb{P}=\mathcal{E}(t) p(x) \psi(x,\vartheta)$.
Here we defined the Boltzmann weight $p(x) \propto e^{-\beta U(x)}$ with $\beta = 1/k_B T$ the inverse temperature, which adopting a convenient normalization reads
\begin{align} \label{eq:peq}
	p(x) = \dfrac{\exp(\beta k x^2/2)}{2 \pi} \; .
\end{align}
The Fokker-Planck operator $\Omega$ in Eq.~\eqref{eq:eom_propagator} appears to be non-Hermitian already in the passive system, $\text{Pe}=0$. 
However, in this case it can be made manifestly Hermitian by a gauge transformation~\cite{Risken1989}.
Here, we circumvent this detour and define a new operator $\mathcal{L}$ by splitting off the Boltzmann weight
\begin{align} \label{eq:splitting}
	\Omega p(x) \psi(x,\vartheta) =: p(x) \mathcal{L} \psi(x,\vartheta) \; .
\end{align}

Inserting into~\eqref{eq:eom_propagator} and using the defined units yields
\begin{align}\label{eq:SE3}
    \frac{1}{\mathcal{E}}\partial_t \mathcal{E} = \dfrac{1}{\psi} \mathcal{L} \psi  \stackrel{!}{=} -\lambda \;,
\end{align}
where the last equal sign holds since the first and second term of the equation are functions of independent variables, and therefore can only be equal to each other if both sides evaluate to a constant.
We can now explicitly write the solution for the time-dependent component of the propagator, $\mathcal{E}(t)=\exp(-\lambda t)$, and proceed to solve the equation for the spatial and angular components only, which now reads
\begin{equation}
    \mathcal{L} \psi + \lambda \psi = 0 \; .
\end{equation}
The similarity to a quantum mechanical problem suggests to tackle the problem by dealing with the activity as a pertubation of the passive system.
We thus split the operator $\mathcal{L}$ into a passive contribution $\mathcal{L}_0$ and an active one $\mathcal{L}_1$ according to
\begin{equation}
    \mathcal{L} = \mathcal{L}_0 + \text{Pe} \, \mathcal{L}_1 \; ,
\end{equation}
with
\begin{gather}
    \mathcal{L}_0 \psi= 
    \dfrac{1}{\tau} \left[ d^2 \partial^2_x + \beta k d^2 x \partial_x  + \gamma \partial_\vartheta^2 \right] \psi \;, \label{eq:L0} \\
    \mathcal{L}_1\psi=-\dfrac{d}{\tau} \left[\beta k x \cos \vartheta + \cos \vartheta \, \partial_x   \right] \psi \; . \label{eq:L1}
\end{gather}

\section*{Solution of the passive reference system}

To solve the unperturbed eigenvalue problem 
\begin{equation} \label{eq:SE4}
    \mathcal{L}_0 \psi = - \lambda \psi \; ,
\end{equation}
subjected to the initial~\eqref{eq:initial_condition} and the boundary~\eqref{eq:boundary_condition_x} conditions, first we decompose the two degree of freedom into different $(n,s)$ modes with the ansatz
\begin{equation} \label{eq:eigenfunctions_ansatz}
    \psi_{n,s}(x,\vartheta)= e^{i s \vartheta} \mathcal{X}_{n}(x) \; .
\end{equation}

\begin{figure}[t!]
\centering
\includegraphics[scale=1]{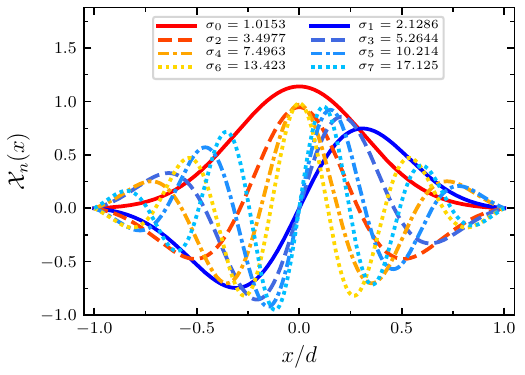} 
\caption{Function $\mathcal{X}_{n}(x) := \exp(-\beta k x^2 /4) Y_n(x) / \mathcal{N}_n$ plotted for $n=0,1,\ldots,7$ with $\beta k d^2 = 10$. Absorbing boundary conditions are imposed in $x=\pm d$ which leads to the reported values of $\sigma_n$ (obtained from numerical calculations).  \label{fig:chi}}
\end{figure}

Inserting Eq.~\eqref{eq:eigenfunctions_ansatz} into~(\ref{eq:SE4}) yields the equation for the $x$ component
\begin{align} \label{eq:Xcomp0}
	\left( \partial^2_x + \beta k x \partial_x    \right) \mathcal{X}_n +   \beta k  \sigma_n   \mathcal{X}_n = 0 \; ,
\end{align}
where 
\begin{equation} \label{eq:EVa}
    \lambda_{n,s}= \dfrac{1}{\tau} \left[ \beta k d^2 \sigma_n +  \gamma s^2 \right] \; .
\end{equation}
Introducing the further ansatz
\begin{align}
	\mathcal{X}_n(x) = e^{-\beta k x^2 / 4} Y_n(x) \; ,
\end{align}
Eq.~\eqref{eq:Xcomp0} becomes
\begin{align}  \label{eq:Xcomp}
	\partial^2_x & Y_n - \left( \dfrac{\beta k}{2} +  \dfrac{\beta^2 k^2 x^2}{4}  - \beta k \sigma_n \right) Y_n  = 0 \; .
\end{align}
The solutions of Eq.~\eqref{eq:Xcomp} are also well known~\cite{Abramowitz1964}.
They are either even or odd functions of $x$, such that we can write~\cite{Caraglio2018}
\begin{align} \label{EQ:Solutions_2}
Y_n (x) \!=\! \!
\left\lbrace \!\!\!
\begin{array}{l} 
 e^{-\beta k x^2 / 4} {_1\!}F_1 \!\! \left(  \dfrac{1}{2} \!-\! \dfrac{\sigma_n}{2} \!; 
\dfrac{1}{2} \!; \dfrac{\beta k x^2}{2} \right) 
\quad n\!=\!0,2,4,\ldots \\
\; \\
\sqrt{\beta k}\, x e^{-\beta k x^2 / 4} \, {_1\!}F_1 \!\! \left( \!  1 \!-\! \dfrac{\sigma_n}{2}  \!; 
\dfrac{3}{2} \!; \dfrac{\beta k x^2}{2} \! \right)
\: n\!=\!1,3,\ldots
\end{array}
\right. 
\end{align}
where $\,_1F_1(a;b;z)$ is the Kummer confluent hypergeometric function.
The boundary conditions $Y_n (\pm d) = 0$ fix the allowed values of $\sigma_n$.
We note the analogy between Eq.~\eqref{eq:Xcomp} and the Schr\"odinger equation for a one-dimensional quantum harmonic oscillator. 
In the ordinary quantum case, the eigenvalues of the Hamiltonian operator are $\propto (n + 1/2)$ with $n=0,1,\ldots$, while eigenfunctios amount to Hermite functions $\propto \exp(-\xi^2/2) H_n(\xi)$, with $\xi$ a proper dimensionless variable and $H_n(\xi)$ the Hermite polynomial of order $n$~\cite{SakuraiQM}.
The latters are obtained by imposing vanishing functions at infinity.
In the present case, imposing vanishing conditions at some finite $x$, as $Y_n (\pm d) = 0$, leads to non-integer values for $\sigma_n$ and to solutions given by the confluent hypergeometric functions~\eqref{EQ:Solutions_2}.
However, since the lowest states are more localized around $x=0$ (see Fig.~\ref{fig:chi}), the $\sigma_n$ for small $n$ are close to the quantum oscillator values $\sigma_n \approx n+1$ (see Fig.~\ref{fig:chi}).
The larger the barrier ($\beta k d^2/2$), the closer is the spectrum of $\sigma_n$ to that of the quantum oscillator.

The explicit expression of the eigenfunctions reads
\begin{align} \label{eq:eigenfunctions}
    \psi_{n,s}(x, & \vartheta)  =  e^{i s \vartheta} \dfrac{e^{-\beta k x^2/4} Y_n(x)}{\mathcal{N}_n} \; ,
\end{align}
where the normalization constant has been chosen such that
\begin{align} \label{eq:normalization}
    \braket{\psi_{n',s'}}{\psi_{n,s}} = \delta_{n,n'} \, \delta_{s,s'} \; .
\end{align}
Here we introduced the Kubo scalar product
\begin{align} \label{eq:KuboScalarProduct}
    \braket{\phi}{\psi} := \int_{-d}^{d} \! \diff x  \int_0^{2\pi} \! \diff \vartheta \, p(x) \phi(x,\vartheta)^* \psi(x,\vartheta) \; ,
\end{align}
and resorted on the fact that the functions $Y_n(x)$ are orthogonal and normalizable with
\begin{align} \label{EQ:Y_normalization}
\int_{-d}^d \!\!  \diff x \, Y_n^2 (x) = \mathcal{N}_n^2 \; .
\end{align}
The isomorphism between $\ket{\psi}$ and $\psi(x,\vartheta)$ is made explicit by introducing generalized position and orientation states $|x \vartheta\rangle$ such that $\psi(x,\vartheta) = \langle x\vartheta |\psi \rangle$.
Using the orthogonality condition~\eqref{eq:normalization} it is easy to see that the operators $\ket{\psi_{n,s}} \bra{\psi_{n,s}}$ are a set of orthogonal projectors and thus we can write the following identity relation
\begin{align} \label{eq:completeness_Hilbert_space}
    \sum_{n,s} \ket{\psi_{n,s}} \bra{\psi_{n,s}} = \mathbb{1} \; ,
\end{align}
where we introduced a compact notation for the summation
\begin{align}
    \sum_{n,s}  :=  \sum_{n=0}^\infty \sum_{s=-\infty}^\infty  \; .
\end{align}

The eigenfunctions of the passive reference system fulfill the completeness relation
\begin{align} \label{eq:completeness}
   p(x) \! \sum_{n,s} \! \psi_{n,s}(x,\vartheta) \psi_{n,s}(x_0,\vartheta_0)^* \!=\!  \delta(x \!-\! x_0) \delta(\vartheta \!-\! \vartheta_0) \; .
\end{align}

The previous completeness relation~\eqref{eq:completeness} allows us to find a solution for the reduced propagator in the passive reference system starting from its formal expression~\eqref{eq:formal_solution}
\begin{align}\label{eq:solution_propagator}
    \mathbb{P}^{(0)} & (x, \vartheta, t | x_0 , \vartheta_0)  =   e^{\Omega t} \delta(x-x_0) \delta(\vartheta-\vartheta_0)   \nonumber \\
    & = p(x) \sum_{n,s} \! \left\lbrace e^{\mathcal{L}_0 t} \psi_{n,s}(x,\vartheta)  \right\rbrace \psi_{n,s}(x_0,\vartheta_0)^* \nonumber \\ 
    & = p(x) \sum_{n,s} \bra{x\vartheta} e^{\mathcal{L}_0 t} \ket{ \psi_{n,s}} \braket{\psi_{n,s}}{x_0 \vartheta_0} \nonumber \\
    & = p(x) \sum_{n,s} e^{-\lambda_{n,s}t} \, \psi_{n,s}(x_0 ,\vartheta_0)^* \, \psi_{n,s}(x ,\vartheta)
\; .
\end{align}
Note that, from the second line of the previous equation and using the identity relation~\eqref{eq:completeness_Hilbert_space}, one can also write
\begin{align}\label{eq:solution_propagator_2}
    \mathbb{P}^{(0)} &(x, \vartheta, t | x_0 , \vartheta_0)  = p(x) \bra{x\vartheta} e^{\mathcal{L}_0 t} \ket{x_0 \vartheta_0}
\; ,
\end{align}
meaning that the propagator is the projection of the generalized position and orientation state $\ket{x\vartheta}$ over the time evolution of the initial state $\ket{x_0\vartheta_0}$, multiplied by the Boltzmann weight $p(x)$.

\section*{Solution for ABP particles}

One readily shows that the passive operator $\mathcal{L}_0$ is Hermitian, $\braket{\phi}{\mathcal{L}_0\psi} = \braket{\mathcal{L}_0 \phi}{\psi}$, with respect to the Kubo scalar product~\eqref{eq:KuboScalarProduct} and consequently its eigenvalues $\lambda_{n,s}$ are real and left and right eigenfunctions coincide, $\ket{\psi_{n,s}^{\text{L}}} = \ket{\psi_{n,s}^{\text{R}}} =\ket{\psi_{n,s}}$.
However, the operator $\mathcal{L}$, taking also care of the particle's activity, does not reflect this property.
Correspondingly, in the following one has to be careful that the eigenvalues of the $\mathcal{L}$ operator, $\lambda_{n,s}^{\text{Pe}}$ are in general complex and the left eigenfunctions, $\ket{\psi_{n,s}^{\text{Pe},\text{L}}}$, are distinct from the right ones $\ket{\psi_{n,s}^{\text{Pe},\text{R}}}$.
If properly normalized, the perturbed left and right eigenfunctions constitute a bi-orthonormal basis with identity relation
\begin{align} \label{eq:completeness_Hilbert_space_perturbed}
    \sum_{n,s} \ket{\psi_{n,s}^{\text{Pe},\text{R}}} \bra{\psi_{n,s}^{\text{Pe},\text{L}}} = \mathbb{1} \; ,
\end{align}
which directly yields the propagator in the presence of activity
\begin{align}\label{eq:solution_propagator_full_problem}
    \mathbb{P} &(x, \vartheta, t | x_0 , \vartheta_0)  =  \bra{x\vartheta} e^{\Omega t} \ket{x_0 \vartheta_0} \nonumber \\
    & = p(x) \sum_{n,s} \bra{x\vartheta} e^{\mathcal{L} t} \ket{ \psi_{n,s}^{\text{Pe},\text{R}}} \braket{\psi_{n,s}^{\text{Pe},\text{L}}}{x_0 \vartheta_0} \nonumber \\
    & = p(x) \sum_{n,s} e^{-\lambda_{n,s}^{\text{Pe}}t} \, \psi_{n,s}^{\text{Pe},\text{L}}(x_0 ,\vartheta_0)^* \, \psi_{n,s}^{\text{Pe},\text{R}}(x ,\vartheta)
\; .
\end{align}

\begin{figure}[t!]
\centering
\includegraphics[scale=1]{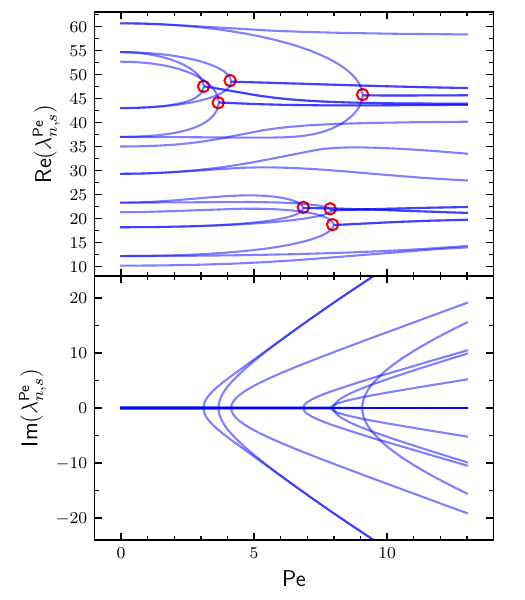} 
\caption{Numerical eigenvalues $\lambda_{n,s}$ of the Fokker-Planck operator $\mathcal{L} = \mathcal{L}_0+\text{Pe} \, \mathcal{L}_1$ as a function of the P{\'e}clet number $\text{Pe}$, for $\beta k d^2 = 10$, $\gamma=2$, $n_{\text{max}}=3$, and $s_{\text{max}}=2$. Transparency of lines and exceptional points highlighted with red circles better show when real components merge and imaginary ones bifurcate.  \label{fig:spectrum}}
\end{figure}

\begin{figure*}[t!]
\centering
\includegraphics[scale=1]{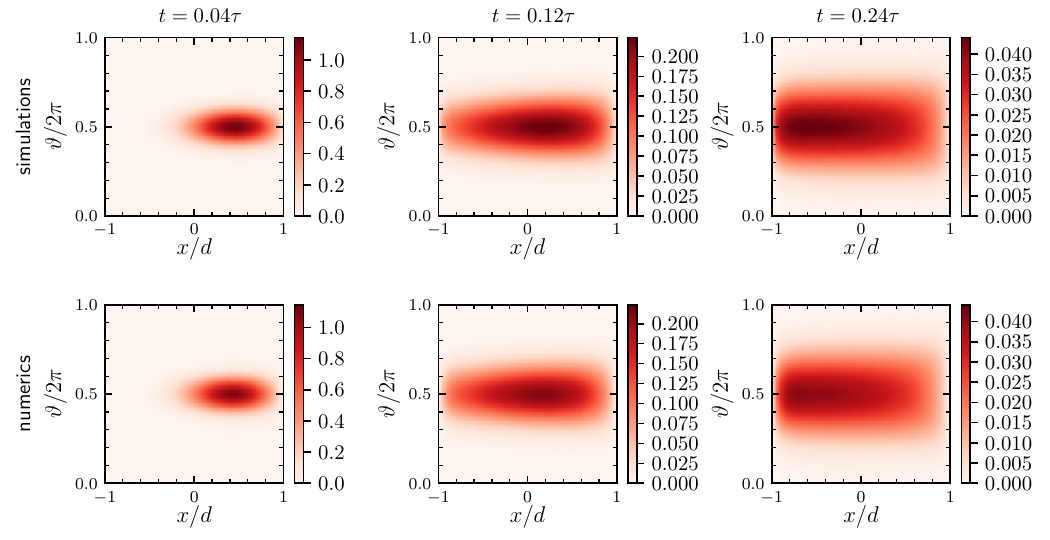}
\caption{Probability distribution at different times $t$ starting with initial condition $x_0 = d/2$ and $\vartheta_0 = \pi$. Comparison between simulations, numerics for $\beta k d^2 =10$, $\text{Pe} = 6$ and $\gamma=2$.
For the simulations, statistics has been collected from $10^8$ independent particles.
For the numerics, $n_{\text{max}}=7$ and $s_{\text{max}}=6$.\label{fig:probability}}
\end{figure*}

\begin{figure}[t!]
\centering
\includegraphics[scale=1]{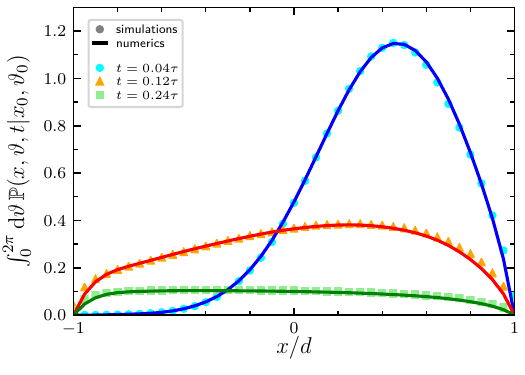}
\caption{Probability distribution, integrated over the self-propulsion direction, $\vartheta$, at different times $t$ starting with initial condition $x_0 = d/2$ and $\vartheta_0 = \pi$. Comparison between simulations, numerics for $\beta k d^2 =10$, $\text{Pe} = 6$ and $\gamma=2$.
For the simulations, statistics has been collected from $10^7$ independent particles.
For the numerics, $n_{\text{max}}=10$ and $s_{\text{max}}=9$.\label{fig:probability_integrated}}
\end{figure}

To explicitly compute the propagator~\eqref{eq:solution_propagator_full_problem}, it is then necessary to calculate the perturbed eigenvalues and left and right eigenfunction.
To this scope, one has first to explicitly evaluate the action of the perturbation $\mathcal{L}_1$ on the eigenstates of $\mathcal{L}_0$.
Starting from Eqs.~\eqref{eq:L1} and~\eqref{eq:eigenfunctions} is it possible to show that
\begin{align}\label{eq:L1action}
\mathcal{L}_1  & \ket{\psi_{n,s}}  \!=\! -\dfrac{d}{2\tau}  \sum_{n'=0}^{\infty} b_{n,n'} \big( \ket{\psi_{n',s+1}} + \ket{\psi_{n',s-1}}\big)  \; ,
\end{align}
with weights $b_{n,n'}$ having a rather lengthy formula which can be found in \hyperref[sec_appA]{appendix A}.

Now, given the finite-dimensional subspace of passive system's eigenfunctions such that $0 \leq n \leq n_{\text{max}}$ and $|s| \leq s_{\text{max}}$, the action of the full operator $\mathcal{L}=\mathcal{L}_0 + \text{Pe} \,\mathcal{L}_1$ is completely characterized by a square matrix $\mathcal{G}$ of dimension $(1+n_{\text{max}}) (2s_{\text{max}} + 1)$ with elements defined by
\begin{align}
& [\mathcal{G}]_{ n' (2s_{\rm max} + 1) + s' +  s_{\rm max} , n  (2s_{\rm max} + 1) + s + s_{\rm max} } \nonumber \\
& \qquad \qquad = \bra{\psi_{n',s'}} \mathcal{L}_0 + \text{Pe} \, \mathcal{L}_1 \ket{\psi_{n,s}} \, ,
\end{align}
which has to be diagonalized numerically to obtain its eigenvalues $\lambda_{n,s}^{\text{Pe}}$ and left and right eigenvectors, $\bra{\psi_{n,s}^{\text{Pe},\text{L}}}$ and $\ket{\psi_{n,s}^{\text{Pe},\text{R}}}$ for any arbitrary P\'eclet number.
The perturbed eigenvectors are then a linear combination of the passive system's eigenstates
\begin{align}
\ket{\psi_{n,s}^{\text{Pe},\text{R}}} & = \sum_{n',s'} g_{n,s}^{\text{R};\, n',s'} \ket{\psi_{n',s'}} \; , 
\label{eq:decomposition_eigenstates1} \\
\bra{\psi_{n,s}^{\text{Pe},\text{L}}} & = \sum_{n',s'} g_{n,s}^{\text{L};\, n',s'} \bra{\psi_{n',s'}} \; . \label{eq:decomposition_eigenstates2}
\end{align}

The computational effort required to diagonalize such matrices increases rapidly with their dimension. 
However, the decaying exponentials in time in the expression of the propagator, Eq.~\eqref{eq:solution_propagator_full_problem}, ensures convergence.
In the unperturbed case, $\text{Pe}=0$, the eigenvalues~\eqref{eq:EVa} are real and an increasing function of $n$ and $|s|$.
However, with increasing P{\'e}clet number, the real components of two distinct eigenvalues may merge and they bifurcate to a pair of complex conjugates for even larger activity, see Fig.~\ref{fig:spectrum}.
These branching points, called exceptional points~\cite{Heiss2012}, often originate in parameter-dependent eigenvalue problems and occur in a great variety of physical problems including mechanics, electromagnetism, atomic and molecular physics, quantum phase transitions, and quantum chaos. 
They have also been observed in other problems concerning active particles~\cite{Kurzthaler2016,Kurzthaler2017,DiTrapani2023}.
The exceptional point are highlighted with red circles in the upper panel of Fig.~\ref{fig:spectrum}.

To corroborate our findings, we benchmark the time evolution of the probability distribution starting from some given initial condition as obtained from numerics against that obtained by direct stochastic simulations, see Fig.~\ref{fig:probability}.
In Fig.~\ref{fig:probability_integrated} we further benchmark numerics against simulation considering the probability distribution integrated over the self-propulsion angle.
Note that due to the presence of the absorbing boundary, the discretization time step adopted in the stochastic simulations should be smaller than what is usually adopted for standard simulations of an ABP in free space.
As a matter of facts, with increasing time, convergence of results in the proximity of the boundary becomes more and more sensitive to the value of the discretization time step, see also appendix B in Ref.~\citep{DiTrapani2023}.

\section*{Transition-Path Times Statistics}

Here, we define transition paths as those trajectories originating at a given point $(x_0=-d+\varepsilon,y_0)$ arbitrarily close to the left boundary ($x=-d$) and being absorbed at any point $(x=d,y)$ on the right boundary in the limit of vanishing $\varepsilon$.

The continuity equation $\partial_t \mathbb{P} = - \vec{\nabla} \cdot \vec{j}$~\cite{note2} allows us to write particle current in the $x$ direction associated to the Fokker–Planck equation
\begin{align}\label{eq:current}
j_x ( x,\! \vartheta  | x_0, \! \vartheta_0)  \!=\!  \dfrac{d^2}{\tau} \left[  \beta k x \!-\! \partial_x \!+\!  \dfrac{\text{Pe}}{d}  \cos \vartheta \right]   \mathbb{P} (x, \! \vartheta | x_0, \! \vartheta_0) \, .
\end{align}
Inserting Eqs.~\eqref{eq:solution_propagator_full_problem},~\eqref{eq:peq},~\eqref{eq:decomposition_eigenstates1}, and~\eqref{eq:decomposition_eigenstates2} into Eq.~\eqref{eq:current} we get
\begin{align}\label{eq:current_2}
 j_x & ( x,\! \vartheta  | x_0, \! \vartheta_0) = \dfrac{d^2}{\tau} p(x) \! \sum_{n,s} \! e^{-\lambda_{n,s}^{\text{Pe}}t} \! \sum_{n',s'} \! g_{n,s}^{\text{L};\, n',s'} \! \psi_{n',s'}(x_0 ,\vartheta_0)^*  \nonumber \\
   &  \times \!\!\! \sum_{n'',s''} \! g_{n,s}^{\text{R};\, n'',s''} \left( \dfrac{\text{Pe}}{d} \cos \vartheta - \partial_x \right) \psi_{n'',s''}(x ,\vartheta)   \ .
\end{align}

The TPT distribution, dependent on the initial angle $\vartheta_0$, is then given by~\cite{Hummer2004,Caraglio2018}
\begin{align}\label{eq_TPTdistr}
& \quad P_{\rm TPT}(t|\vartheta_0) = \lim_{\varepsilon \to 0} \dfrac{  \int_0^{2\pi} \! \diff \vartheta \, j_x (d,\vartheta,t | -d+\varepsilon, \vartheta_0)}{ \int_0^{\infty} \!\! \diff t'  \int_0^{2\pi} \! \diff \vartheta \, j_x (d,\vartheta,t' | -d+\varepsilon, \vartheta_0) } \, .
\end{align}

Recalling that the boundary condition imposes $\psi(\pm d,\vartheta)=0$, we have that
\begin{align}
\psi(-d+\varepsilon,\vartheta) \approx \varepsilon \partial_x \psi(-d,\vartheta)\, ,
\end{align}
and that the current at the $x=d$ boundary appearing in Eq.~\eqref{eq_TPTdistr} does not show an explicit dependence on the P{\'e}clet number.
Note, however, that such a current is still inderectly depending on the activity since the $g^{\rm L}$'s and $g^{\rm R}$'s coefficients change when vaying the P{\'e}clet number.
We can thus write
\begin{align}
& j_x (d,\vartheta,t | -d+\varepsilon, \vartheta_0) \approx - \varepsilon \dfrac{d^2}{\tau} p(d) \sum_{n,s} e^{-\lambda_{n,s}^{\text{Pe}}t} \nonumber \\
   & \quad \times \! \sum_{n',s'}  g_{n,s}^{\text{L};\, n',s'} \, \partial_x \, \psi_{n',s'}(-d, \vartheta_0)^*  \nonumber \\
   & \quad \times \! \sum_{n'',s''}  g_{n,s}^{\text{R};\, n'',s''} \partial_x  \psi_{n'',s''}(d,\vartheta)   \ .
\end{align}
This current is of order $\varepsilon$, however this factor is canceled by using the normalization constant in Eq.~\eqref{eq_TPTdistr}, and hence the TPT distribution in the limit $\varepsilon \rightarrow 0$ remains finite. 
Integrating also over $\vartheta$,one finally obtains
\begin{align}\label{eq_TPTdistr_final}
& \quad P_{\rm TPT}(t|\vartheta_0) = \dfrac{\displaystyle \sum_{n,s} e^{-\lambda_{n,s}^{\text{Pe}}t} G^{\text{L}}_{n,s}(-d, \vartheta_0) \, G^{\text{R}}_{n,s}(d)}
{\displaystyle \sum_{n,s} \dfrac{1}{\lambda_{n,s}^{\text{Pe}}}  G^{\text{L}}_{n,s}(-d, \vartheta_0) \, G^{\text{R}}_{n,s}(d)  } \, ,
\end{align}
with
\begin{align}
G^{\text{L}}_{n,s}(x, \vartheta) :=  \sum_{n',s'}  g_{n,s}^{\text{L};\, n',s'} e^{-i s' \vartheta}  \partial_x \mathcal{X}_{n'}(x) \, ,
\end{align}
and
\begin{align}
G^{\text{R}}_{n,s}(x) :=  \sum_{n'}   g_{n,s}^{\text{R};\, n',0} \, \partial_x \mathcal{X}_{n'}(x)  \, ,
\end{align}
with $\mathcal{X}_{n}(x) := \exp(-\beta k x^2 /4) Y_n(x) / \mathcal{N}_n$.
In the limit of vanishing activity ($\text{Pe}=0$) one recovers the TPT distribution of a one-dimensional passive particle crossing a parabolic barrier as reported in Ref.~\cite{Caraglio2018}.

\begin{figure}[t!]
\centering
\includegraphics[scale=1]{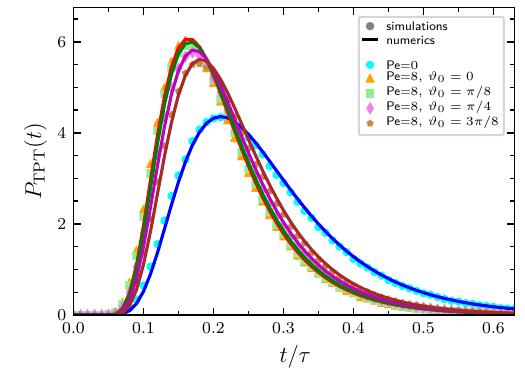}
\caption{Transition-path-time distribution, $P_{\rm TPT}(t) = P_{\rm TPT}(t| \vartheta_0)$, as a function of the direction of the self-propulsion at the left boundary, $\vartheta_0 = 0$.
Comparison between simulations (symbols) and numerics (lines) for $\beta k d^2 =10$ and $\gamma=2$
For the simulations, statistics has been collected from $10^6$ independent transition paths defined as paths starting at $x=-d+\varepsilon$ with $\varepsilon = 10^{-7}$ and ending at $x=d$.
For the numerics, $n_{\rm max}=32$ and $s_{\rm max}=6$.
\label{fig:tpt_distro}}
\end{figure}

Our results show that the peak of the TPT distribution of an active particle for a relatively high barrier ($\beta k d^2 = 10$) is shifted to the left with respect to the peak of the TPT distribution of a passive particle, see Fig.~\ref{fig:tpt_distro}.
Surprisingly, the TPT distribution depends only mildly on the initial direction of the self-propulsion when starting from the left boundary ($x=-d$).
Furthermore, we also note that the average TPT decreases with the activity of the particle. 
Also the coefficient of variation~\cite{Satija2020}, which is a useful measure of the distribution broadness, decreases with the activity, being about $0.41$ for the passive case in Fig.~\ref{fig:tpt_distro} and displaying a slightly smaller value for the active cases. 
Interestingly, this result aligns with what is observed in one-dimensional systems~\cite{Carlon2018} while the opposite behavior is observed in different two-dimensional systems~\cite{Zanovello2021,Zanovello2021b}.

\section*{Survival probability and first-passage-time distribution}

The knowledge of the reduced propagator allows computing also the survival probability at time $t$.
The latter, given some initial conditions $(x_0, \vartheta_0)$, is readily obtained by integrating over the final $x$ coordinate and orientation
\begin{align}\label{eq:survival_probability}
S(t|x_0, \vartheta_0) \! = \! \int_{-d}^d \!\! \diff x  \int_0^{2\pi} \!\!\!\! \diff \vartheta \, \mathbb{P}(x,\vartheta,t|x_0,\vartheta_0) \; .
\end{align}
Since 
\begin{align}
\int_{-d}^d \!\! \diff x  \int_0^{2\pi} \!\!\!\! \diff \vartheta  \, p(x) \psi_{n,s}(x,\vartheta) = \delta_{s,0} \, f_n  \; ,
\end{align}
with
\begin{align}
f_{n} \! := \! \left\lbrace  \!\!\!
\begin{array}{l}
\dfrac{1}{\mathcal{N}_n} \displaystyle \! \int_{-d}^d \!\!\! \diff x \; {_1\!}F_1 \!\! \left(  \dfrac{1}{2} \!-\! \dfrac{\sigma_n}{2} \!; 
\dfrac{1}{2} \!; \dfrac{\beta k x^2}{2} \right)  \mbox{  if } n=0,2,\ldots \\
\\
 0  \qquad \qquad \mbox{else} \, ,
\end{array}
\right. 
\end{align}
exploiting Eqs.~\eqref{eq:solution_propagator_full_problem},~\eqref{eq:decomposition_eigenstates1}, and~\eqref{eq:decomposition_eigenstates2}, one obtains
\begin{align}\label{eq:survival_probability2}
   S&(t|x_0, \vartheta_0)  = \sum_{n,s} e^{-\lambda_{n,s}^{\text{Pe}}t} \nonumber \\
   & \times \!\!\! \sum_{n',s'}  g_{n,s}^{\text{L};\, n',s'} \psi_{n',s'}(x_0 ,\vartheta_0)^*  \sum_{n''}  g_{n,s}^{\text{R};\, n'',0}  f_{n''} \, .
\end{align}

\begin{figure}[t!]
\centering
\includegraphics[scale=1]{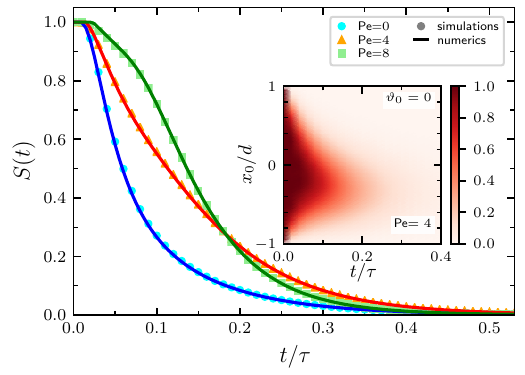}
\caption{Survival probability, $S(t) = S(t| x_0,\vartheta_0)$, as a function of time for different $\text{Pe}$ and with initial condition $x_0=-d/2$ and $\vartheta_0 = 0$.
Comparison between simulations (symbols) and numerics (lines) for $\beta k d^2 =10$ and $\gamma=0.4$
For the simulations, statistics has been collected from $10^5$ independent particles.
For the numerics, $n_{\rm max}=12$ and $s_{\rm max}=6$.
Inset: Survival probability as a function of time and $x_0$ for $\text{Pe}=4$ as obtained from numerics for the remaining initial condition $\vartheta_0 = 0$.
\label{fig:survival_probability}}
\end{figure}

See Fig.~\ref{fig:survival_probability} for a comparison between the results at different P{\'e}clet numbers obtained by numerics and by direct stochastic simulations.
As expected, due to their activity and the persistence of their motion, active particles display a different behavior with respect to standard passive Brownian particles.
For example, starting with an initial position at $x_0=-d/2$ we can note that a passive particle has a $50\%$ probability of being absorbed at the boundaries within a time of about $0.06\tau$, see Fig.~\ref{fig:survival_probability}.
This quick initial decay of the survival probability is due to the fact that the passive particle is likely unable to overcome the potential barrier and is soon absorbed at the left boundary of the box ($x=0$).
On the other hand, with increasing activity and initial direction pointing towards the barrier ($\vartheta_0=0$), an ABP particle has more and more chances of crossing the barrier which results in an initially slower decay of the survival probability.
However, once the active particle reaches the peak of the energy potential, the self-propulsion starts enhancing the probability of being absorbed at longer times.
Consequently, the survival probability decreases faster at longer times with increasing activity, see Fig.~\ref{fig:survival_probability}.
In fact, when plotting the survival probability as a function of the initial position $x=0$ for an initial self-propulsion direction pointing towards the right boundary ($\vartheta_0=0$), we note an asymmetric shape with a faster decay for $x_0 > 0$ and the slowest decay observed for values of $x_0$ in the range $(-d/2, -d/5)$, see inset of Fig.~\ref{fig:survival_probability}.
In contrast, a passive particle would show a symmetric behavior with the slowest decay centered at $x_0=0$.

\begin{figure}[t!]
\centering
\includegraphics[scale=1]{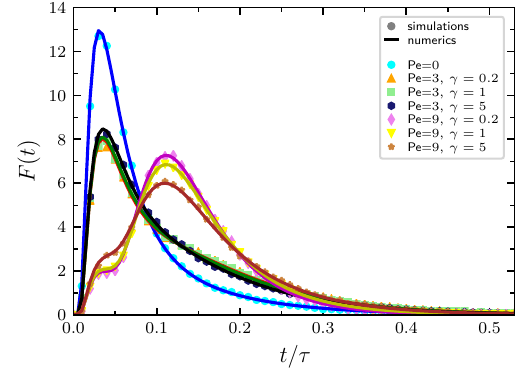}
\caption{First-passage-time distribution, $F(t) = F(t| x_0,\vartheta_0)$ for different $\text{Pe}$ and $\gamma$ and with initial condition $x_0=-d/2$ and $\vartheta_0 = 0$.
Comparison between simulations (symbols) and numerics (lines) for $\beta k d^2 =10$.
For the simulations, statistics has been collected from $10^7$ independent particles.
For the numerics, $n_{\rm max}=60$ and $s_{\rm max}=5$.
\label{fig:fpt}}
\end{figure}

Furthermore, starting from Eq.~\eqref{eq:survival_probability} we can also obtain the first-passage-time distribution for any given initial condition as
\begin{align}\label{eq:first_passage_def}
F(t|x_0, \vartheta_0) = - \dfrac{\diff S(t|x_0, \vartheta_0)}{\diff t} \; .
\end{align}
As for the survival probability, the ABP exhibits first-passage properties that differ from those of a passive particle.
In particular, again starting with the initial state ($x_0=-d/2$, $\vartheta_0=0$), the first-passage-time distribution at large P{\'e}clet numbers shows a bump at short times and a faster decay at longer times, see Fig.~\ref{fig:fpt}.
One can also note that, at least for the considered initial conditions and parameters, the rotational diffusivity influences the shape of distribution only to a limited extent, see Fig.~\ref{fig:fpt}.
Intuitively, when the persistence length is larger or comparable to the distance separating the two absorbing boundaries, our results are comparable with those obtained in the case of a single absorbing wall~\cite{Bauche2025}.

Note that to reach convergence in the first-passage-time distribution a much larger value of $n_{\rm max}$ is needed in comparison to that used to obtain the survival probability or the probability distribution.
This observation is due to the fact that, when taking the time derivative of the survival probability, a factor equal to the eigenvalue $\lambda_{n,s}^{\text{Pe}}$ appears in the series expansion of the first-passage-time distribution.
Thus, while already for the survival probability it is necessary to consider more and more terms in the summation~\eqref{eq:survival_probability2} when time decreases, this issue becomes even more important for the first-passage-time distribution.

\section*{Conclusions}

We have derived an exact series solution for the probability propagator (along the $x$ coordinate and the self-propulsion angle) of a two-dimensional ABP navigating in a domain with absorbing boundaries at $x=\pm d$ and subjected to a parabolic potential along the $x$ direction, centered at $x=0$.
Such a solution is possible by taking the standard passive Brownian motion as a reference system and dealing with the activity of the particle as a perturbation.
However, differently from the approximate solutions usually achieved with perturbation theory~\cite{SakuraiQM}, our solution is exact.
The reduced propagator is expressed in terms of the left and right eigenvectors, which can be easily computed by direct diagonalization of the matrix form of the Fokker-Planck operator, multiplied by an exponentially decaying factor with a rate given by the corresponding perturbed eigenvalue.
Note that, to make the numerics possible, we have to cut off the number of considered eigenfunctions but convergence is ensured by the decaying exponentials in time in the expression of the propagator, Eq.~(\ref{eq:solution_propagator_full_problem}).
The approach adopted in this work takes inspiration from the one recently adopted to solve the ABP model in a circular harmonic trap~\cite{Caraglio2022}.
However, in contrast to what happens in the case of an isotropic harmonic potential, in the present case switching on the particle activity induces a change in the eigenvalue spectrum of the Fokker-Planck operator.
With increasing activity more and more eigenvalues become complex and exceptional points~\cite{Heiss2012} arise.

The knowledge of the propagator is then exploited as a starting point to obtain several observables.
In particular, given some initial condition, we compute the spatial probability density at a later time (integration of the propagator over the self-propulsion orientation $\vartheta$), the survival probability (integration over all coordinates), and the related first-passage-time distribution.
All these quantities show a strong dependence on the activity of the particle and, to a lesser extent, on its rotational diffusivity.
This observation is in line with what observed for and ABP exploring a circular domain with an absorbing boundary~\cite{DiTrapani2023}.
We also derive the distribution of the transition-path time, i.e. of the time needed to travel from the boundary on one side of the barrier to the boundary on the other side.

Besides extending the nowadays limited set of exactly solvable models for active particles~\cite{Wagner2017,Hermann2018,Schnitzer1993,Tailleur2008, Tailleur2009,Malakar2018,Kurzthaler2016,Kurzthaler2017,Kurzthaler2018,Martens2012,Sevilla2015,Caraglio2022,DiTrapani2023}, our findings also pile on the recent literature regarding first-passage properties of active particles~\cite{Malakar2018,Angelani2014,Scacchi2018,Demaerel2018,Dhar2019,Basu2018,Caprini2021} and can be exploited to reach analytical insight into target-search~\cite{Tejedor2012,Volpe2017,Zanovello2021,Zanovello2021b} problems in more complex environments involving absorbing boundaries.
Furthermore, an analytical expression for TPT distribution in the case of active forces have been previously obtained only in the case of one-dimensional systems~\cite{Carlon2018}.
Thus, our derivation of the TPT distribution in the presence of absorbing boundaries for a two-dimensional active particle represents a significative advance in the field.
Finally, our conceptual strategy of solving the Fokker-Planck equation of an ABP model in two-dimension by disregarding one of the two spatial coordinate may also be adopted in other problems as for example an ABP in presence of an absorbing wall or in a periodic potential modulated along one direction.

\begin{acknowledgments}
This research was funded in whole by the Austrian Science Fund (FWF) [10.5576/P35872]. M.C. acknowledges Thomas Franosch and Enrico Carlon for fruitful discussions. In memory of Carlo Vanderzande, who essentially contributed to the research reported in Ref.~\cite{Caraglio2018} and consequently also inspired the present one.
\end{acknowledgments}
\medskip

\appendix

\begin{widetext}

\section{Action of $\mathcal{L}_1$ on the passive system's eigenfunction $\ket{\psi_{n,s}}$} \label{sec_appA}
Here, we now explicitly evaluate the action of the perturbation $\mathcal{L}_1$ on the eigenstates of $\mathcal{L}_0$ obtained from the eigenvalue problem
\begin{align}
    \mathcal{L}_0 \ket{\psi_{n,s}} = -\lambda_{n,s} \ket{\psi_{n,s}} \; .
\end{align}
Considering Eqs.~\eqref{eq:L1} and~\eqref{eq:eigenfunctions} we have
\begin{align} \label{eq:L1action_App}
\mathcal{L}_1  & \ket{\psi_{n,s}}  = -\dfrac{d}{\tau} \Bigg\{  \left( \dfrac{e^{i(s+1)\vartheta} + e^{i(s-1)\vartheta}}{2} \right) (\beta k x + \partial_x) \dfrac{e^{-\beta k x^2/4} Y_n(x)}{\mathcal{N}_n}  \Bigg\} \; ,
\end{align}
and
\begin{align} \label{eq:L1action_x}
 (\beta k x + \partial_x) \dfrac{e^{-\beta k x^2/4} Y_n(x)}{\mathcal{N}_n} = \left\lbrace 
\begin{array}{ll} 
\dfrac{\beta k(1\!-\!\sigma_n)}{\mathcal{N}_n}  x   e^{-\beta k x^2 / 2} {_1\!}F_1 \! \left(  \dfrac{3}{2} \!-\! \dfrac{\sigma_n}{2} ; \dfrac{3}{2} ; \dfrac{\beta k x^2}{2} \right)  & \quad n\!=\!0,2,4,\ldots \\
 & \\ 
 \dfrac{1}{\mathcal{N}_n} \sqrt{\beta k} e^{-\beta k x^2 / 2} \, {_1\!}F_1 \! \left( \!  1 \!-\! \dfrac{\sigma_n}{2}  ;  \dfrac{3}{2} ; \dfrac{\beta k x^2}{2} \! \right) &  \\
 \quad + \dfrac{2-\sigma_n}{3 \mathcal{N}_n} (\beta k)^{3/2} x^2  e^{-\beta k x^2 / 2} {_1\!}F_1 \! \left(  2 \!-\! \dfrac{\sigma_n}{2} ; \dfrac{5}{2} ; \dfrac{\beta k x^2}{2} \right) & \quad n\!=\!1,3,\ldots
\end{array}
\right. 
\end{align}
where we used the property~\cite{NIST}
\begin{align}
 \partial_{z} \; {_1\!}F_1 (a;b;z) = \dfrac{a}{b} \; {_1\!}F_1 (a+1;b+1;z) \; .
\end{align}
These function are odd for $n=0,2,4,\ldots$ and even for $n=1,3,\ldots$.

We note then that given a complete orthogonal system of functions $\{ \phi_{\ell}(z) \}$ over the interval $\mathcal{R}$, the functions  $\phi_{\ell}(z)$ satisfy an orthogonality relationship of the form 
\begin{align} \label{eq:general_orthogonality}
\int_{\mathcal{R}} \diff z\, w(z) \phi_{\ell'}(z) \phi_\ell(z)   = c_\ell \delta_{\ell,\ell'} \; ,
\end{align}
where $w(z)$ is a a weighting function, $c_\ell$ are given constants and $\delta_{\ell,\ell'}$ is the Kronecker delta. 
An arbitrary function $f(z)$ can be written as a series
\begin{align} \label{eq:general_series}
f(z)= \sum_{\ell=0}^{\infty} a_\ell \phi_\ell(z) \; , 
\end{align}
with
\begin{align} \label{eq:general_coefficients}
a_\ell = \dfrac{1}{c_\ell} \int_{\mathcal{R}} \diff z \, w(z)  \phi_{\ell}(z) f(z)  \; . 
\end{align}

Substituting $z$ with $x$ and applying relations~\eqref{eq:general_orthogonality}-\eqref{eq:general_coefficients} to the case in which $\phi_n(x) = e^{-\beta k x^2/4} Y_n(x) / \mathcal{N}_n$, $f(x)$ is given by Eq.~\eqref{eq:L1action_x}, $w(x) = e^{\beta k x^2/2}$, and $c_{n}=1$ we have
\begin{align} \label{eq:L1action_x_2}
& \quad (\beta k x + \partial_x) \dfrac{e^{-\beta k x^2/4} Y_n(x)}{\mathcal{N}_n} = \! \sum_{n'= 0}^{\infty} b_{n,n'} \dfrac{e^{-\beta k x^2/4} Y_{n'}(x)}{ \mathcal{N}_{n'}} \; ,
\end{align}
with
\begin{align}
 b_{n,n'} = \left\lbrace 
 \begin{array}{ll}
\dfrac{\beta k(1\!-\!\sigma_n)}{\mathcal{N}_n \mathcal{N}_{n'}}  \displaystyle \int_{-d}^d \diff x \, e^{-\beta k x^2/4} Y_{n'}(x) \, x \, {_1\!}F_1 \! \left(  \dfrac{3}{2} \!-\! \dfrac{\sigma_n}{2} ; \dfrac{3}{2} ; \dfrac{\beta k x^2}{2} \right)
    & \mbox{ if } n=0,2,\ldots \mbox{ and } n'=1,3,\ldots \\ 
    & \\
\dfrac{\sqrt{\beta k}}{\mathcal{N}_n \mathcal{N}_{n'}} \displaystyle \int_{-d}^d \diff x \, e^{-\beta k x^2/4} Y_{n'}(x)   \Bigg[ {_1\!}F_1 \! \left( \!  1 \!-\! \dfrac{\sigma_n}{2}  ;  \dfrac{3}{2} ; \dfrac{\beta k x^2}{2} \! \right) & \\
\quad + \dfrac{2-\sigma_n}{3} \beta k x^2 {_1\!}F_1 \! \left(  2 \!-\! \dfrac{\sigma_n}{2} ; \dfrac{5}{2} ; \dfrac{\beta k x^2}{2} \right) \Bigg]
    & \mbox{ if } n=1,3,\ldots \mbox{ and } n'=0,2,\ldots \\
    & \\
 0  & \mbox{ otherwise.} \\
 \end{array}
 \right. 
\end{align}

Finally, inserting the above relations into Eq.~\eqref{eq:L1action_App} one gets
\begin{align} \label{eq:L1action_App_2}
\mathcal{L}_1  & \ket{\psi_{n,s}}  = -\dfrac{d}{2\tau}  \sum_{n'=0}^{\infty} b_{n,n'} \big( \ket{\psi_{n',s+1}} + \ket{\psi_{n',s-1}}\big) \; .
\end{align}
\end{widetext}


\begin{thebibliography}{78}
\bibitem [{not()}]{note2}%
  \BibitemOpen
  \href@noop {} {}\bibinfo {note} {Note that here we abused notation by writing
  $\vec{\nabla}=(\partial_x,\partial_\vartheta)$ and $\vec{j}=(j_x,
  j_\vartheta)$, thus including in the vectorial notation a $\vartheta$
  component.}\BibitemShut {Stop}%
\end{thebibliography}
\end{document}